\documentclass[preprint,english,superscriptaddress, floatfix,longbibliography]{revtex4-2}
\usepackage{amsmath}
\usepackage{graphicx}
\usepackage{CJK}
\usepackage{amsgen}
\usepackage{amsfonts}
\usepackage{amsbsy}
\usepackage{amssymb}
\usepackage{dcolumn}
\usepackage{bm}
\usepackage{epsfig}
\usepackage{color}
\usepackage{lineno}
\usepackage{booktabs}
\usepackage{orcidlink}

\begin{document}

\title{General recipe for immediate entanglement death and birth via Bell
states: Environmental Heisenberg exchange with transition as an example}

%Can qubit states be pushed through the entanglement-unentanglement Boundaries?%
%Can Qubit States Cross the Entanglement-Unentanglement Divide%

%\preprint{}


\begin{CJK*}{UTF8}{bsmi}

\author{Son-Hsien Chen (陳松賢)\thanks{*Corresponding Author}~\orcidlink{0000-0002-3700-0018}}
\email{sonhsien@utaipei.edu.tw}
\affiliation{Department of Applied Physics and Chemistry, University of Taipei, Taipei 100234, Taiwan}
\author{Seng Ghee Tan (陳繩義)~\orcidlink{0000-0002-3233-9207}}
\email{csy16@ulive.pccu.edu.tw}
\affiliation{Department of Optoelectric Physics, Chinese Culture University, Taipei 11114, Taiwan}
\author{Che-Chun Huang (黃則鈞)}
\email{f06222009@ntu.edu.tw}
\affiliation{Division of Advanced Metrology, Taiwan Semiconductor Manufacturing Company Limited, Hsinchu 300091, Taiwan}

%\author{}
%\thanks{}
%\affiliation{}


\date{\today}

\begin{abstract}
Environment is known to play a dual role in both extinguishing and establishing entanglement, leading to entanglement sudden death (ESD) and entanglement sudden birth (ESB). In this paper, we propose a recipe for the initial states of two qubits to undergo ESD, ESB, or transition of finite duration (TFD) between them. The recipe is explained through a physical picture of entanglement penetrability. While this recipe is \emph{general} (i.e., it does not require a specific type of interaction), a spin-star model with environmental Heisenberg exchange is chosen for illustration. We introduce the entanglement switch parameter (ESP), whose sign indicates whether bipartite entanglement between the qubits is switched on or off. Utilizing Bell states, we show how the penetrability of the ESP enables the observation of ESD and ESB. The classical (quantum) weighting of the Bell states encodes the ESP for initial mixed (pure) states. When more than two Bell states are adopted, the ESP permits states to penetrate through the entanglement-unentanglement boundary. In this case, the penetrability of a small ESP ensures the immediate occurrence of ESD or ESB and indicates the TFD if the local time-even symmetry in the entanglement monotone is also satisfied. When no more than two Bell states are employed, the penetrability is lost, and TFD is only identified in some mixed states but not in pure states. Thanks to the simplicity of this model, analytic results are provided. We also analyze the symmetries that can convert or alter ESD into ESB, and vice versa. The recipe enhances the controllability of entanglement dynamics and facilitates entanglement engineering.
\end{abstract}

\keywords{entanglement, quantum science, Heisenberg model, spintronics, magnetism}

\maketitle

\end{CJK*}

\section{Introduction}

\label{sec:intro} Quantum entanglement~\cite{Horodecki2009}, which
characterizes constituent subsystems that are not tensor-product separable,
plays a crucial role in quantum technologies~\cite{Bennett2000}, including
gravitational wave detection~\cite{Khalili2018,Zeuthen2019}, quantum
cryptography~\cite{Ekert1991,Shor2000,Gisin2002, Bennett1998}, and quantum
computation~\cite{Horodecki2009,DiVincenzo1995,Steane1998}, such as error correction and scalable architectures~\cite{DiVincenzo2000,Ladd2010,Nielsen2010,Madsen2022}.
In addition, entanglement
underpins a wide range of essential protocols in quantum information
processing, such as quantum key distribution\cite%
{Bennett1992,Kwek2021,Li2022,Liu2022}, quantum secure direct
communication\cite{Long2002,Deng2003,Zhou2022}, controlled quantum
teleportation \cite{Karlsson1998,Deng2005,Jing2024}, remote state preparation%
\cite{Nawaz2018,Lan2024}, and entanglement purification\cite%
{Bennett1996,Yan2023}. Once established, this quantum correlation can
persist even when each subsystem is isolated~\cite{Chen2024}. On the other
hand, in open systems where an interacting environment or medium is
introduced to the subsystems, the environment triggers decoherence~\cite%
{Schlosshauer2019} and can play dramatically different roles---inducing
entanglement, suppressing it, or even both.

In the case of suppression, interactions with environment can lead to the
disappearance of entanglement. However, unlike typical subsystem decoherence
process, where coherence dies out asymptotically~\cite{Vicari2018,Arzano2023}%
, the alteration from entanglement to unentanglement can occur suddenly for
both mixed and pure initial states~\cite{Yashodamma2013}. For instance, a
pioneering study predicted that two entangled atoms, each with two levels
and encapsulated in their cavities, can abruptly become unentangled
after a finite time due to vacuum fluctuations via spontaneous emissions~%
\cite{Yu2004a}. This phenomenon, known as entanglement sudden death (ESD)~%
\cite{Yu2006a,Yu2006b,Ann2007,Yu2009}, was later confirmed experimentally
using optical setups~\cite{Almeida2007}. Furthermore, it was recently
pointed out~\cite{Chen2024} that ESD is more likely to arise in the minimal
set of bipartite entanglement, consisting of a pair ($N=2$) of qubits ($d=2$%
), but less likely in the entanglement beyond the minimal set, $N>2$ or $d>2$%
. Here, $d$ indicates the qudit dimension, i.e., the number of eigenstates
(or energy levels) a subsystem can occupy, and $N$ specifies the number of
qudits. Indeed, most existing research identifying ESD pertains to the
minimal set~\cite{AlQasimi2008,Yu2009,Yashodamma2013,Sharma2020},
in spin systems~\cite{Wang2018,Chen2024}, optical systems~\cite{Almeida2007},
and other systems under different noisy~\cite{Yu2004a,Yashodamma2013}
and non-Markovian~\cite{Bellomo2007,Bellomo2008} environments. The ESD has also been experimentally observed~in solid-state systems~\cite%
{Wang2018}. A few techniques have been developed to postpone ESD~\cite%
{Rau2008,Qian2012,Chakraborty2019,Awasthi2023}. By adjusting the initial
states within certain amount of entanglement, excitation, and purity, the
ESD can be avoided~\cite{Qian2012}.

It is also well known that subsystem
entanglement can be provoked~\cite{Schneider2002,Benatti2003,Hartmann2020}
by interacting with thermal~\cite{Braun2002,Kim2002a,Valido2013}, atomic~\cite{Chen2022},
and optical environments~\cite{HorMeyll2009,Passos2018}.
During measurement, the environment interacts with subsystems, allowing
quantum information to be sensed and extracted; even random interactions can
establish, preserve, and extend entanglement~\cite{Kong2020}, as well as
alter the spatial scaling law~\cite{Lunt2020,Turkeshi2021,Botzung2021} of
the saturated entanglement. In addition, entanglement mechanism has garnered
significant attention and is considered essential in the field of
spintronics~\cite{Yuan2022}. In particular, recent developments employing
inelastic neutron scattering ~\cite{Mathew2020,Laurell2021,Scheie2021} and
X-ray scattering~\cite{Suresh2024,Hales2023,Liu2025} have enabled the
detection of entanglement in spintronics devices. Notably, two distant
macroscopic ferromagnets can become entangled through spin exchange mediated
by electrons~\cite{Suresh2024}, disrupting the conservation of local
ferromagnet magnetization. This conservation can be process-dependent,
influenced by whether the dynamics is non-Markovian or Markovian~\cite%
{GarciaGaitan2024}. The resulting non-conservation of magnetization limits
the applicability of classical spin equations of motion, such as the
Landau-Lifshitz-Gilbert equation. The mediated entanglement is also realized
in a similar microscopic model known as the spin-star network~\cite%
{Hutton2004}, where all subsystem spins interact exclusively with a central
environmental spin through Heisenberg exchange. In this model, the
entanglement mediating capability of the central spin depends on the total
number of subsystem spins. Moreover, in the minimal set of two qubits~\cite%
{Qiang2009,He2018,Mostafa2022}, not only has ESD been identified, but the
revival of entanglement, conventionally termed entanglement sudden birth
(ESB), has also been discovered~\cite{Yuan2007}.

In light of the dual role of environments in both inducing and suppressing
entanglement, this paper examines the transition between ESD and ESB [Fig.~%
\ref{fig:sche}(a)]. This transition, often recognized as entanglement
rebirth, has been explored in various contexts~\cite{Ficek2006,Ficek2008,Das2009},
including spin~\cite{Qiang2009} and atomic systems~\cite{Tanas2010}; however, a common (interaction-independent) strategy to accomplish such a transition has been lacking. In this study, we
introduce the entanglement switch parameter (ESP) $\varepsilon $ based on
Bell states. The ESP is defined so that its sign indicates whether the
entanglement is switched on or off. By assuming that a sufficiently small
ESP does not affect the overall entanglement tendency but only slightly
displaces the trajectories in the Hilbert space, we present a general recipe
for preparing initial states that yield the immediate onset of ESD or ESB.
If the \emph{additional} condition that the local time-even symmetry in the
entanglement monotones is satisfied, a nontrivial (namely, finite)
transition time can be achieved. This transition of finite duration (TFD)
emerges in both pure and mixed states. Without loss of generality, we choose
to illustrate the recipe using a simple model---perhaps the simplest---the
spin-star Heisenberg model with two subsystems [Fig.~\ref{fig:sche}(b)].
Each subsystem contains a spin-1/2 qubit, with the magnitude of the central
spin effectively accounting for the environmental degrees of freedom. The
simplicity of the model allows the arguments to be presented transparently,
and even analytic results can be obtained. Nonetheless, other spin-product
forms, such as the Dzyaloshinskii-Moriya interaction~\cite%
{Qiang2009,Qiang2010}, remain applicable. The findings herein provide
guidelines for manipulating and controlling entanglement, offering a
practical and systematic method that can be effectively adopted in designing
and advancing quantum technology applications.

It should be clarified that the present work differs from studies on typical tripartite entangled states, such as energy-time-entangled GHZ and W states~\cite{MacLean2018,Li2022a,Li2024}, which mainly focus on static correlations in closed systems. A recent monogamy-based study~\cite{Mannai2025} showed that when one qubit acts as an environment, the bipartite entanglement in GHZ states vanishes, while it remains in W states. However, since this disentanglement in GHZ states requires a specific equal-weight configuration, it results in a zero-duration transition. In contrast, the aim of the present work is to provide a general recipe by which ESD and ESB can occur over a \emph{finite} time interval through environment-induced dynamics.

The paper is organized as follows. Section~\ref{sec:mot} describes the
studied model and reviews the entanglement monotones. The results are
discussed in Sec.~\ref{sec:rl}. We first examine the symmetries relevant to
the occurrence of, and connections between, ESD, ESB, and the death-birth
transition in Sec.~\ref{sec:rl_sya}, where we also show that the TFD is
rare. Section~\ref{sec:rl_rec} then outlines the recipe for achieving the
transition utilizing initial Bell states with ESPs. The transition is
inspected for both mixed states in Sec.~\ref{sec:rl_cla} and pure states in
Sec.~\ref{sec:rl_qua}. Section~\ref{sec:summ} summaries the results.

\section{Model and entanglement monotone}

\label{sec:mot}
\begin{figure}[tbph]
%\begin{figure*} for two-column figure
\centerline{\psfig{file=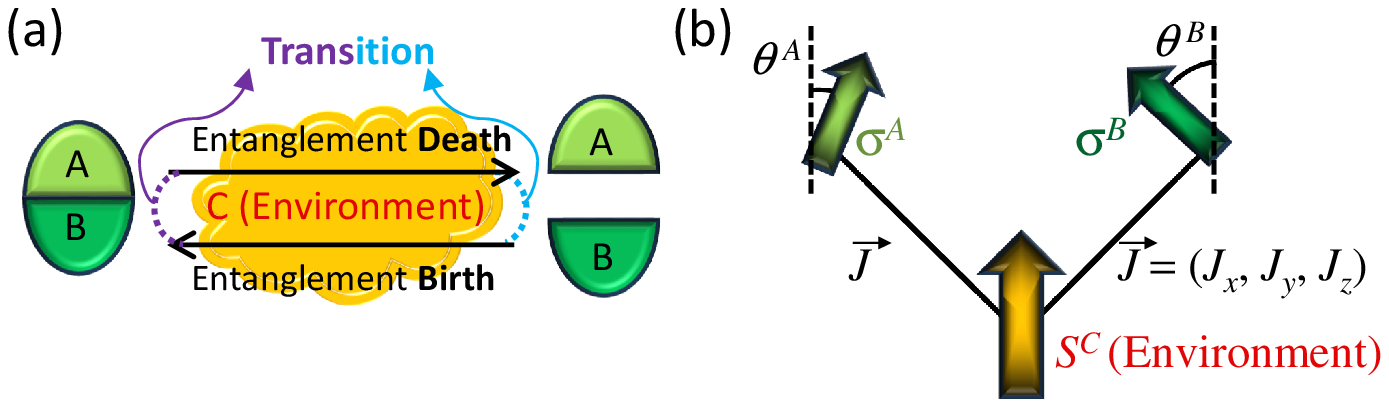,scale=0.68,angle=90,clip=true,
trim=130mm 23mm 0mm 0mm}}%For angle=90(270), trim=b(t),r(l),t(b),l(r)
%t=top,b=bottom,r=right,l=left
%Below is for inserting another figure.
%\centerline{\psfig{file=figX.eps,scale=0.33,angle=270,clip=true,trim=0mm 0mm 0mm 0mm}}
\caption{(a) Schematic of the environment-$C$-induced bipartite $A$-$B$
entanglement dynamics, including entanglement death (right arrow),
entanglement birth (left arrow), transition from death to birth (blue dotted
line), and transition from birth to death (purple dotted line). (b)
Spin-star model illustrating two spin-1/2 subsystems, $\protect\sigma ^{A}$
and $\protect\sigma ^{B}$, both coupled to a spin-$S^{C}$ environment
through Heisenberg exchange interactions $\vec{J}$ (solid lines). The
conventional spherical coordinate system is used with spin-$z$ axis at $%
\left( \protect\theta ,\protect\phi \right) =\left( 0,0\right) $; for
example, the spin directions of $A$ and $B$ are represented by the polar
angles $\protect\theta _{A}$ and $\protect\theta _{B}$, respectively.}
\label{fig:sche}
\end{figure}

To realize ESD, ESB, and the transition between them in Fig. \ref{fig:sche}%
(a), consider the spin-star model in Fig. \ref{fig:sche}(b). This model
consists of two spin-1/2 qubits, $A$ and $B$, interacting with a common
environment $C$ through Heisenberg exchange coupling $\vec{J}=\left(
J_{x},J_{y},J_{z}\right) $. The environment carrying quantum degrees of
freedom with spin $S^{C}$, can be either pure $Tr\left[ \left( \rho
^{C}\right) ^{2}\right] =1$ or mixed $Tr\left[ \left( \rho ^{C}\right) ^{2}%
\right] <1$, where $\rho ^{C}$ represents the reduced density matrix (DM)
for $C$. The Hamiltonian reads,%
\begin{equation}
H_{\vec{J}}=\sum_{\alpha =x,y,z}J_{\alpha }S_{\alpha }^{C}\sigma _{\alpha
}^{A}+J_{\alpha }S_{\alpha }^{C}\sigma _{\alpha }^{B}  \label{eq:Hami}
\end{equation}%
in which $\sigma _{\alpha }^{A}$ and $\sigma _{\alpha }^{B}$ are Pauli
matrices of the qubit subsystems, and $S_{\alpha }^{C}$ is the spin-$S^{C}$
matrix in direction $\alpha \in \left\{ x,y,z\right\} $, e.g., $S_{\alpha
}^{C}=\sigma _{\alpha }^{C}/2$ if $S^{C}=1/2$. The reduced Planck's constant
$\hbar $ is set to $\hbar =1$ here. Note that the model (\ref{eq:Hami}) can
be experimentally carried out in quantum dots using a mediated idle spin~%
\cite{Chan2021}. The time evolution operator $U\left( \vec{J},t\right) $
governs the dynamics of the total wave function $\left\vert \psi \left(
t\right) \right\rangle =U\left( \vec{J},t\right) \left\vert \psi
_{0}\right\rangle =\left\vert S^{C},\sigma ^{A},\sigma ^{B}\right\rangle
\left( t\right) $, with%
\begin{equation}
U\left( \vec{J},t\right) =\exp \left( -iH_{\vec{J}}t\right) \text{,}
\label{eq:U_Jt}
\end{equation}%
rendering the equation of motion for the DM,%
\begin{equation}
\frac{d}{dt}\varrho \left( t\right) =-i\left[ H_{\vec{J}},\varrho \left(
t\right) \right] \text{,}  \label{eq:emo_rho}
\end{equation}%
where the commutator is defined as $\left[ X,Y\right] \equiv XY-YX$. The
initial state can be prepared with classical weights $w_{i}$ via
\begin{eqnarray}
\varrho \left( t=0\right) &=&\varrho _{0}  \notag \\
&=&\sum_{i}w_{i}\left\vert \psi _{i}\right\rangle \left\langle \psi
_{i}\right\vert \text{,}  \label{eq:Wweight}
\end{eqnarray}%
while with quantum weights $c_{i}=\sqrt{w_{i}}$ over the orthonormal basis $%
\left\vert \chi _{i}\right\rangle $ via
\begin{equation}
\left\vert \psi _{0}\right\rangle =\sum_{i}c_{i}\left\vert \chi
_{i}\right\rangle \text{.}  \label{eq:Qweight}
\end{equation}%
From Eq. (\ref{eq:emo_rho}), the short-time (immediate) behavior of the
system after interacting with the environment is then harnessed by the
leading terms of the expansion,%
\begin{equation}
\varrho \left( t\right) =\varrho _{0}+\left( -i\right) \left[ H_{\vec{J}%
},\varrho _{0}\right] dt+\frac{1}{2!}\left( -i\right) ^{2}\left[ H_{\vec{J}},%
\left[ H_{\vec{J}},\varrho _{0}\right] \right] dt^{2}+O\left( dt^{3}\right)
\text{.}  \label{eq:dtexp}
\end{equation}

As depicted in Fig. \ref{fig:sche}(a), we are interested in the dynamics of
bipartite entanglements between $A$ and $B$, whose corresponding reduced DM
reads,%
\begin{equation}
Tr_{C}\left[ \varrho \left( t\right) \right] =\rho ^{AB}\left( t\right)
\equiv \rho \left( t\right)
\end{equation}%
in which environment $C$ is traced out. One of the entanglement monotones,
which faithfully~\cite{Bhaskara2017,Nourmandipour2021} reflects the
two-qubit entanglement, is the negativity~\cite{Vidal2002}%
\begin{equation}
\mathcal{N}(\vec{J},t)=Tr\sqrt{\rho ^{T_{B}}\left( t\right) ^{\dagger }\rho
^{T_{B}}\left( t\right) }\text{,}  \label{eq:N_Jt}
\end{equation}%
computed as the absolute sum of all negative eigenvalues of the partial
transpose%
\begin{equation}
\left[ \rho ^{T_{B}}\left( t\right) \right] _{ab,a^{\prime }b^{\prime
}}=\rho _{ab^{\prime },a^{\prime }b}\left( t\right) \text{.}
\label{eq:PTB_rho}
\end{equation}%
Here subscripts $a$ and $a^{\prime }$ ($b$ and $b^{\prime }$) denote the
indices belonging to $A$ ($B$). The negativity $Tr\sqrt{\rho ^{T_{A}}\left(
t\right) ^{\dagger }\rho ^{T_{A}}\left( t\right) }=Tr\sqrt{\rho
^{T_{B}}\left( t\right) ^{\dagger }\rho ^{T_{B}}\left( t\right) }$ is
irrelevant to which subsystem is transposed. In the case of two qubits,
another equivalent entanglement monotone, called the concurrence \cite%
{Hill1997,Wootters1998,Rungta2001} $\mathcal{C}$ can also be used to
quantify entanglement by finding the maximum,%
\begin{equation}
\mathcal{C}(\vec{J},t)=\max \left\{ 0,2\gamma _{\max }-\Gamma \right\}
\label{eq:C_Jt}
\end{equation}%
with $\gamma \in \left\{ \gamma _{1},\gamma _{2},\gamma _{3},\gamma
_{4}\right\} $ being the eigenvalues of the matrix $\sqrt{\sqrt{\rho \left(
t\right) }\rho ^{\prime }\left( t\right) \sqrt{\rho \left( t\right) }}$, $%
\gamma _{\max }${}$=\max \left( \gamma \right) $, and $\Gamma =\left( \gamma
_{1}+\gamma _{2}+\gamma _{3}+\gamma _{4}\right) $. Here $\rho ^{\prime
}\left( t\right) $ is constructed, from the complex conjugate $\rho ^{\ast
}\left( t\right) $ of the four-by-four reduced DM $\rho \left( t\right) $
and the tensor product $\sigma _{y}^{\otimes 2}=\sigma _{y}\otimes \sigma
_{y}$, as
\begin{equation}
\rho ^{\prime }\left( t\right) =\sigma _{y}^{\otimes 2}\rho \left( t\right)
^{\ast }\sigma _{y}^{\otimes 2}.
\end{equation}%
Being worth mentioning, the evolved density matrix, $\varrho \left( t\right)
$ and reduced DM, $\rho \left( t\right) $, $\rho ^{T_{A}}\left( t\right) $,
and $\rho ^{T_{B}}\left( t\right) $, depend on the interaction $\vec{J}$,
whereas the initial DM $\varrho _{0}$ and the reduced bipartite $\rho _{0}$
do not. As we will see later, the immediacy to encounter ESD and ESB can be
controlled primarily by the interaction-independent ESP $\varepsilon $
solely encoded in the initial conditions.

\section{Result and discussion}

\label{sec:rl}

We begin with analyzing how entanglement responds to different symmetries.
Even in the special case of direct $A$-$B$ exchange,
\begin{equation}
H_{\vec{J}}^{dir}=\sum_{\alpha =x,y,z}J_{\alpha }^{dir}\sigma _{\alpha
}^{A}\sigma _{\alpha }^{B}
\end{equation}%
we gain valuable insights into how the anisotropy of exchange $\vec{J}$
induces entanglement. Using an expansion similar to Eq. (\ref{eq:dtexp}),
the immediate (near-future $dt\approx 0$) concurrence can be computed
straightforwardly as,%
\begin{equation}
\mathcal{C}^{dir}(\vec{J}^{dir},dt)=2\left\vert dt\left(
J_{y}^{dir}-J_{x}^{dir}\cos \theta _{B}\right) \right\vert +O\left(
dt^{2}\right) \text{,}  \label{eq:Cdir}
\end{equation}%
with defining $\sigma ^{A}$ in spin-$z$, $\theta _{A}\equiv 0$, and $\theta
_{B}$ being the angle between the two initial unentangled spin directions.
It is interesting to note that the immediate entanglement does not depend on
the out-of-plane $J_{z}^{dir}$ but solely on the in-plane $J_{x}^{dir}$ and $%
J_{y}^{dir}$. If two initial spins are both in spin-$z$, $\theta _{B}=0$,
then isotropic $J_{x}^{dir}=J_{y}^{dir}$ does not elicit any entanglement.
However, if in-plane exchange is anisotropic (i.e., spin configuration of
lowest energy prefers some special in-plane directions), entanglement $%
\mathcal{C}^{dir}(\vec{J}^{dir},dt)>0$ emerges. If $\sigma ^{A}$ and $\sigma
^{B}$ are not collinear, $\vec{n}^{A}\times \vec{n}^{B}\neq 0$, then
entanglement can be easily evoked as indicated by the coordinate-free form
of Eq. (\ref{eq:Cdir}),
\begin{equation}
\mathcal{C}^{dir}(\vec{J}^{dir},dt)\approx 2\left\vert dt\right\vert \times
\left\vert \vec{J}^{dir}\cdot \left( \vec{n}^{A}\times \vec{n}^{B}\right) +%
\vec{J}^{dir}\cdot \left[ \vec{n}^{A}\times \left( \vec{n}^{A}\times \vec{n}%
^{B}\right) \right] \left( \vec{n}^{A}\cdot \vec{n}^{B}\right) \right\vert
\label{eq:Cdir_free}
\end{equation}%
where $\vec{n}^{A/B}$ is the unit vector in the initial spin $\sigma ^{A/B}$
direction. Furthermore, within the small interval $\left[ -dt,dt\right] $,
the perfector $\left\vert dt\right\vert $ in the monotone (\ref{eq:Cdir_free}%
) implies the local (around $t\approx 0$) time-even symmetry, $\mathcal{C}%
^{dir}(\vec{J}^{dir},dt)=\mathcal{C}^{dir}(\vec{J}^{dir},-dt)$. While in the
above case the entanglement death-birth transition is trivial, i.e., of zero
duration, as we will see later, this symmetry is instrumental in
constructing the immediate TFD using the ESP.

\begin{table}[tbph]
\caption{Early development of the characteristic negative eigenvalue $%
\protect\lambda ^{\ast }(dt)$ for different fully unentangled and individually pure initial
spin configurations $\left\vert S^{C},\protect\sigma ^{A},\protect\sigma %
^{B}\right\rangle $, as a function of the environment spin $S^{C}$. The
highest spin-$z$ eigenstate of $C$ is represented by $\left\vert \uparrow ,%
\protect\sigma ^{A},\protect\sigma ^{B}\right\rangle =\left\vert m=S^{C},%
\protect\sigma ^{A},\protect\sigma ^{B}\right\rangle $.}
\label{tab:sepABC}\centering
%Put tabular Here
\begin{tabular}{c|c}
\toprule$%
\begin{array}{c}
\text{Configuration} \\
\left\vert S^{C},\sigma ^{A},\sigma ^{B}\right\rangle%
\end{array}%
$ & $%
\begin{array}{c}
\text{Characteristic} \\
\text{negative eigenvalue }\lambda ^{\ast }\left( dt\right)%
\end{array}%
$ \\ \hline
$\left\vert \uparrow ,\uparrow ,\uparrow \right\rangle $ & $%
\begin{array}{c}
dt^{2}S^{C}\times \left\{
\begin{array}{cc}
J_{y}\left( J_{y}-J_{x}\right) \text{,} & \text{if }\left\vert
J_{x}\right\vert >\left\vert J_{y}\right\vert \\
J_{x}\left( J_{x}-J_{y}\right) \text{,} & \text{if }\left\vert
J_{x}\right\vert <\left\vert J_{y}\right\vert%
\end{array}%
\right. \\
+O\left( dt^{3}\right) \text{, sign}\left( J_{x}\right) =\text{sign}\left(
J_{y}\right)%
\end{array}%
$ \\ \hline
$\left\vert \uparrow ,\uparrow ,\downarrow \right\rangle $ & $%
\begin{array}{c}
dt^{2}S^{C}\left( J_{x}^{2}+J_{y}^{2}-\sqrt{\left(
J_{x}^{2}+J_{y}^{2}\right) ^{2}+4J_{x}^{2}J_{y}^{2}}\right) \\
+O\left( dt^{3}\right)%
\end{array}%
$ \\ \hline
$\left\vert \uparrow ,\downarrow ,\downarrow \right\rangle $ & $%
\begin{array}{c}
dt^{2}S^{C}\times \left\{
\begin{array}{cc}
J_{y}\left( J_{x}+J_{y}\right) \text{,} & \text{if }\left\vert
J_{x}\right\vert >\left\vert J_{y}\right\vert \\
J_{x}\left( J_{x}+J_{y}\right) \text{,} & \text{if }\left\vert
J_{x}\right\vert <\left\vert J_{y}\right\vert%
\end{array}%
\right. \\
+O\left( dt^{3}\right) \text{, sign}\left( J_{x}\right) =-\text{sign}\left(
J_{y}\right)%
\end{array}%
$ \\
\bottomrule
\end{tabular}%
\end{table}

\begin{figure*}[tbph]
%\begin{figure*} for two-column figure
\centerline{\psfig{file=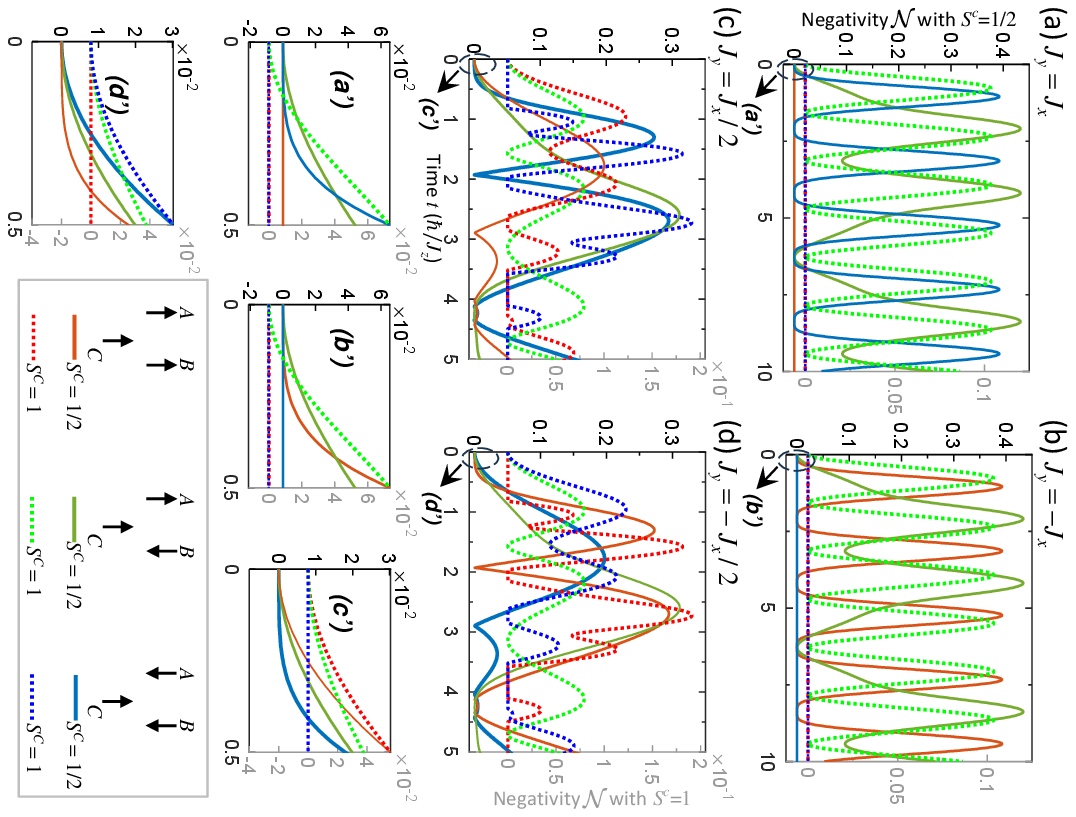,scale=0.95,angle=90,clip=true,
trim=18mm 120mm 0mm 0mm}}%For angle=90(270), trim=b(t),r(l),t(b),l(r)
%t=top,b=bottom,r=right,l=left
%Below is for inserting another figure.
%\centerline{\psfig{file=figX.eps,scale=0.33,angle=270,clip=true,trim=0mm 0mm 0mm 0mm}}
\caption{Evolution of entanglement negativity, $\mathcal{N}(\vec{J},t)$, for
the initial unentangled spin states $\left\vert S^{C},\protect\sigma ^{A},%
\protect\sigma ^{B}\right\rangle =\left\vert \uparrow ,\uparrow ,\uparrow
\right\rangle $ (red), $\left\vert \uparrow ,\uparrow ,\downarrow
\right\rangle $ (green), and $\left\vert \uparrow ,\downarrow ,\downarrow
\right\rangle $ (blue) with $S^{C}=1/2$ (solid lines, left axis) and $%
S^{C}=1 $ (dotted lines, right axis), at different exchange coupling, (a) $%
\vec{J}=(1,1,1)$, (b) $\vec{J}=(1,-1,1)$, (c) $\vec{J}=(1,0.5,1)$, and (d) $%
\vec{J}=(1,-0.5,1)$, in units of $J_{z}\equiv 1$. Corresponding zoomed-in
views around $t=0$, consistent with the expressions in Table \protect\ref%
{tab:sepABC}, are shown in (a'), (b'), (c'), and (d'). Note that the blue
(red) solid line in (a) [(b)] shows that the negativity initially grows at
an order higher than $dt^{2}$ before oscillating periodically, i.e., no
finite-duration transitions are found for all cases of $S^{C}=1/2$ within
the chosen time window.}
\label{fig:longt}
\end{figure*}

Now, let us return to the environment-mediated case (\ref{eq:Hami}) and
focus on the negativity (\ref{eq:N_Jt}) by characterizing the eigenvalues.

\paragraph*{\textbf{Definition 1.}}

The characteristic negative eigenvalue (CNE) $\lambda ^{\ast }$ is defined
as the most-negative (or the smallest) eigenvalues in the partial transpose
of the DM $\rho ^{T_{A}/T_{B}}$.

Unless otherwise specified, throughout this paper, for all initial states
studied in the limit $dt\rightarrow 0^{+}$, we find there is \emph{at most}
one negative eigenvalue in Eq.~(\ref{eq:PTB_rho}), which pinpoints the CNE $%
\lambda ^{\ast }\left( dt\right) $. This eigenvalue determines whether the
system exhibits immediate negativity,
\begin{equation}
\mathcal{N}(\vec{J},dt)=\max \left\{ 0,-\lambda ^{\ast }\left( dt\right)
\right\} .
\end{equation}%
For different $S^{C}$ and for fully unentangled and individually pure initial
states $\left\vert S^{C},\sigma ^{A},\sigma ^{B}\right\rangle =\left\vert
\uparrow ,\uparrow ,\uparrow \right\rangle $, $\left\vert \uparrow ,\uparrow
,\downarrow \right\rangle $, and $\left\vert \uparrow ,\downarrow
,\downarrow \right\rangle $, using expansion (\ref{eq:dtexp}), we compute
the CNE in Table~\ref{tab:sepABC}. The $\left\vert \uparrow ,\sigma
^{A},\sigma ^{B}\right\rangle =\left\vert m=S^{C},\sigma ^{A},\sigma
^{B}\right\rangle $ denotes $C$ in its highest spin-$z$ eigenstate. The
corresponding negativity dynamics is shown in Fig. \ref{fig:longt}, where
exchange $J_{z}=1$ as the energy unit, and thus $\hbar /J_{z}$ as the time
unit is adopted. Although the physical exchange interactions discovered so far are limited (see, for example, Table I in Ref.~\cite%
{Hazzard2014}), Fig.~\ref{fig:longt} suggests that ESD and ESB can be induced for various types of $\vec{J}$. However, for all $S^{C}=1/2$,
represented by the solid lines, the transition durations are trivial within
the chosen time window; that is, no TFD is obtained. The zero durations are
also verified by examining the concurrence $\mathcal{C}(\vec{J},t)$ (not
shown here). Specifically, the $\left\vert \uparrow ,\uparrow ,\uparrow
\right\rangle $ blue solid line in Fig.~\ref{fig:longt}(a) [the $\left\vert
\uparrow ,\downarrow ,\downarrow \right\rangle $ red solid lines in Fig.~\ref%
{fig:longt}(b)] indicates that the negativity initially develops at an order
higher than $dt^{2}$, as inferred from Table~\ref{tab:sepABC}, and then
undergoes periodic oscillation. In addition, Fig.~\ref{fig:longt} shows that
when an opposite sign is assigned in the in-plane exchange, $%
J_{y}\rightarrow -J_{y}$, the environmental spin field is effectively
reversed, turning the negativity of the initial state from $\left\vert
\uparrow ,\uparrow ,\uparrow \right\rangle $ (blue lines) into that of $%
\left\vert \uparrow ,\downarrow ,\downarrow \right\rangle $ (red lines).
Although no immediate TFD is seen in Fig.~\ref{fig:longt}, the TFD can still
be found after a sufficient wait; for instance, see the blue and red dotted
lines ($S^{C}$ $=1$) around $t=4$ in Figs. \ref{fig:longt}(c) and~\ref%
{fig:longt}(d). One can thus monitor the entanglement until the TFD emerges
and identify the states at the onset of the transition. However, these
states, along with the transition beginning time and the transition duration
identified through this monitoring process, are case-dependent and difficult
to control. As will be seen later, a systematic method for identifying the
states leading to TFD can be established. Before addressing this method, it
is worth discussing how symmetry can be applied to relate ESD and ESB in
systems with the same or different $\vec{J}$.

\subsection{Symmetry analysis}

\label{sec:rl_sya}

It is evident from~Eq. (\ref{eq:U_Jt}) that the parameter $t$ can be
absorbed into $\vec{J}$ in the spin-star model (\ref{eq:Hami}) as $U\left(
\vec{J},t\right) =U\left( \vec{J}t,1\right) $, leading to
\begin{equation}
U\left( \vec{J},-t\right) =U\left( -\vec{J},t\right) \text{.}
\label{eq:sys_UJmt_UmJt}
\end{equation}%
Hence, the effect of the swap $\vec{J}\rightarrow -\vec{J}$ is equivalent to
running the time backward. As a consequence, the existence of ESD (ESB) in
the $\vec{J}$ environmental exchange reflects the presence of ESB (ESD) in $-%
\vec{J}$. The death-to-birth (birth-to-death) transition in
antiferromagnetic $\vec{J}$ guarantees the birth-to-death (death-to-birth)
transition in ferromagnetic $-\vec{J}$.

Another similar dynamic reversal can be argued based on the time-reversal
symmetry, which our current system possesses. Let $\mathcal{T}$ be the
time-reversal operator, and consider a system that obeys the time-reversal
symmetry $\mathcal{T}H(t)\mathcal{T}^{\dagger }=H(t)$, with $H(t)=H_{\vec{J}%
} $ in our case. Applying $\mathcal{T}$ to the equation of motion (\ref%
{eq:emo_rho}) yields,
\begin{equation}
\frac{d}{d\bar{t}}\bar{\varrho}(\bar{t})=i[H(\bar{t}),\bar{\varrho}(\bar{t})]%
\text{,}  \label{eq:sys_drhobar_dtbar}
\end{equation}%
with $\bar{t}\equiv -t$. This means that the same equation of motion in a
\emph{rewound} form of (\ref{eq:emo_rho}) governs $\bar{\varrho}(\bar{t})$.
Accordingly, at any given time $t$, by preparing the initial state $\bar{%
\varrho}(\bar{t})=\mathcal{T}\varrho (t)\mathcal{T}^{\dagger }$---which
corresponds to flipping all the spins in the eigenspinors that decompose $%
\varrho (t)$---one can recover the original initial negativity $\mathcal{N}%
(t=0)$. That is, if the initial state $\varrho _{0}$ evolves at time $t$
after encountering ESD (ESB), then the final-as-initial reversal state $%
\mathcal{T}\varrho (t)\mathcal{T}^{\dagger }$ will visit ESB (ESD) in the
same system.

Furthermore, the symmetry that resides in the vicinity of the
entanglement-unentanglement boundary is the local time-even symmetry,%
\begin{equation}
\mathcal{N}(\vec{J},dt)=\mathcal{N}(\vec{J},-dt)\text{,}
\label{eq:sys_NJt_NJmdt}
\end{equation}%
or simply $dt^{2}$ symmetry as referred to herein. This symmetry is
characterized by the $dt$-dependent leading order $dt^{2n}$, with $%
n=1,2,\cdots $, of the monotone. Unlike time-reversal symmetry, the $dt^{2}$
symmetry depends on the initial state. We elaborate further on the
dependence of negativity on $dt$ below. Note that a partial trace over $C$
must be performed to obtain the reduced DM $\rho $. Since $\rho $ is a
four-by-four matrix, expanding to the second term in (\ref{eq:dtexp})
contributes to $\mathcal{N}(\vec{J},dt)$ up to the fourth order, $dt^{4}$,
while the third term contributes at orders $dt^{2}$, $dt^{4}$, $dt^{6}$, and
$dt^{8}$. Therefore, it is sufficient to show that $\mathcal{N}(\vec{J},dt)$
is at least of order $dt^{2}$ if the expansion up to the second term in Eq. (%
\ref{eq:dtexp}) does not yield any linear $dt$ term in $\mathcal{N}(\vec{J}%
,dt)$.

\begin{figure*}[tbph]
%\begin{figure*} for two-column figure
\centerline{\psfig{file=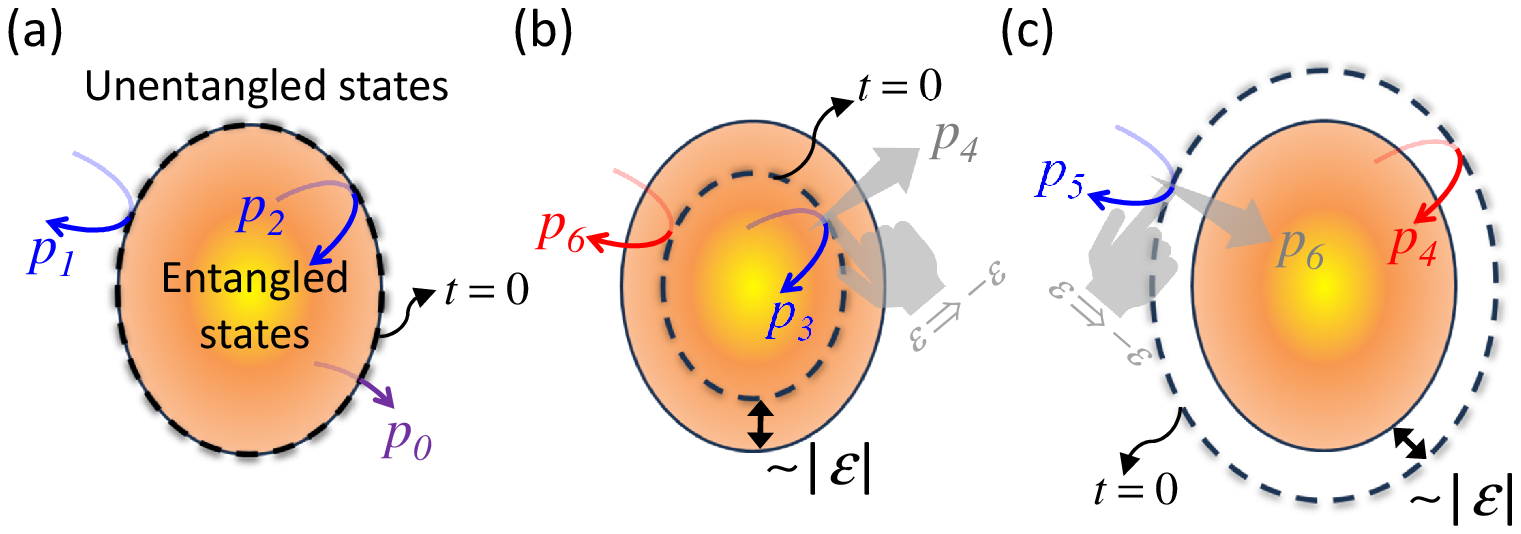,scale=0.63,angle=90,clip=true,
trim=112mm 3mm 0mm 0mm}}%For angle=90(270), trim=b(t),r(l),t(b),l(r)
%t=top,b=bottom,r=right,l=left
%Below is for inserting another figure.
%\centerline{\psfig{file=figX.eps,scale=0.33,angle=270,clip=true,trim=0mm 0mm 0mm 0mm}}
\caption{Schematic in the Hilbert space for the trajectories, $p_{0}$, $%
p_{1} $, $\cdots $, and $p_{6}$, approaching the boundary between entangled
(colored inner region) and unentangled (colorless outer region) states near
the initial time $t=0$ (dashed lines). (a) $t=0$ is set at the boundary, (b)
$t=0$ is set slightly toward the entangled region, quantified by $|\protect%
\epsilon |$, and (c) $t=0$ is set slightly toward the unentangled region,
also quantified by $|\protect\epsilon |$. The entanglement switch parameter $%
\protect\epsilon $ is penetrable if it has opposite signs in (b) and (c),
such that $\protect\epsilon \rightarrow -\protect\epsilon $ shifts $p_{3}$
to $p_{4}$ and $p_{5}$ to $p_{6}$ (refer to the schematic of the gray
index-finger-dragging arrows indicating the shift direction). For
trajectories $p_{1,2,...,6}$, the local time-even symmetry in entanglement
monotones allows the validity of the entire trajectory. However, if the
system lacks this symmetry, half of the trajectory of $t>0$ remains
applicable.}
\label{fig:traj}
\end{figure*}

The $dt^{2}$ symmetry corresponds to the trajectories $p_{1}$ or $p_{2}$
approaching the entanglement-unentanglement boundary in the Hilbert space,
as shown schematically in Fig.~\ref{fig:traj}. The trajectories $p_{1}$ and $%
p_{2}$ in Fig. \ref{fig:traj} imply trivial transition. The nontrivial
transition with ESD or ESB around $t=0$, requires the monotone to have a
leading $dt$-dependence term of order $dt^{2n-1}$. For example, $p_{0}$ in
Fig.~\ref{fig:traj} indicates the occurrence of ESD. However, $dt^{2}$
symmetry actually appears in a wide variety of initial states. Table~\ref%
{tab:sepABC} already shows that $dt^{2}$ symmetry manifests in fully
unentangled and individually pure initial states. In fact, unentangled mixed initial
states of the form,
\begin{equation}
\varrho _{0}=\rho ^{C}\otimes \left\vert \sigma ^{A}\right\rangle
\left\langle \sigma ^{A}\right\vert \otimes \left\vert \sigma
^{B}\right\rangle \left\langle \sigma ^{B}\right\vert \text{,}
\label{eq:rho0_CABspe}
\end{equation}%
also exhibit $dt^{2}$ symmetry. Here $\rho ^{C}$ is in diagonal form,%
\begin{equation}
\rho ^{C}=\left(
\begin{array}{ccccc}
\rho _{S^{C}}^{C} &  &  &  & 0 \\
& \rho _{S^{C}-1}^{C} &  &  &  \\
&  & \ddots  &  &  \\
&  &  & \rho _{-S^{C}+1}^{C} &  \\
0 &  &  &  & \rho _{-S^{C}}^{C}%
\end{array}%
\right) \text{.}
\end{equation}%
Denoting the initial spin directions of $\sigma ^{A/B}$ by $\left( \theta
^{A/B},\phi ^{A/B}\right) $ in spherical coordinates, and taking terms \emph{%
up to the second} in (\ref{eq:dtexp}), the CNE for initial states (\ref%
{eq:rho0_CABspe}) can be computed as,%
\begin{equation}
\lambda ^{\ast }=dt^{2}\frac{J_{z}^{2}}{2}\times \left(
\sum_{m=-S^{C},-S^{C}+1,\cdots ,S^{C}-1,S^{C}}m\rho _{m}^{C}\right)
^{2}\times \left( -2+\cos 2\theta _{A}+\cos 2\theta _{B}\right) +O\left(
dt^{2}\right) \text{.}  \label{eq:lamda_star_Csepmix}
\end{equation}%
Note that the remaining $O\left( dt^{2}\right) $ above originates from the
third term in expansion (\ref{eq:dtexp}).

The $dt^{2}$ symmetry also shows up in the $A$-$B$ entangled pure initial
states of the form, $\left\vert \psi _{0}\right\rangle =\left\vert \uparrow
\right\rangle \otimes \left\vert \alpha _{p}^{+}\right\rangle ,$ $\left\vert
\uparrow \right\rangle \otimes \left\vert \alpha _{p}^{-}\right\rangle $, $%
\left\vert \uparrow \right\rangle \otimes \left\vert \beta
_{p}^{+}\right\rangle $, and $\left\vert \uparrow \right\rangle \otimes
\left\vert \beta _{p}^{-}\right\rangle $. Here,%
\begin{equation}
\left\vert \alpha _{p}^{\pm }\right\rangle =\sqrt{\frac{1+p}{2}}\left\vert
\uparrow ,\uparrow \right\rangle \pm \sqrt{\frac{1-p}{2}}\left\vert \uparrow
,\uparrow \right\rangle
\end{equation}%
and%
\begin{equation}
\left\vert \beta _{p}^{\pm }\right\rangle =\sqrt{\frac{1+p}{2}}\left\vert
\uparrow ,\downarrow \right\rangle \pm \sqrt{\frac{1-p}{2}}\left\vert
\uparrow ,\downarrow \right\rangle
\end{equation}%
are the partially ($0<p<1$) or fully ($p=0$) entangled Bell states. Again,
considering \emph{up to the second term} in (\ref{eq:dtexp}), the CNE for
different $S^{C}$ varies with $dt^{2}$ as,%
\begin{equation}
\lambda ^{\ast }=\frac{-\sqrt{1-p^{2}}}{2}\left[ 1+8\left(
S^{C}J_{z}dt\right) ^{2}\right] +O\left( dt^{2}\right)
\label{eq:lambda_star_alpha}
\end{equation}%
for $\left\vert \alpha _{p}^{\pm }\right\rangle $ and%
\begin{equation}
\lambda ^{\ast }=\frac{-\sqrt{1-p^{2}}}{2}+O\left( dt^{2}\right)
\label{eq:lambda_star_beta}
\end{equation}%
for $\left\vert \beta _{p}^{\pm }\right\rangle $. Therefore, the
entanglement remains for small $dt$, namely, no immediate ESD. At first
glance, it appears that $dt^{2}$ symmetry exists in abundant initial
conditions as shown above but does not lead to any ESD, ESB, and TFD.
However, if a parameter $\varepsilon $ is introduced to \emph{shift the
symmetry point} of the $dt^{2}$ symmetry slightly away from $dt=0$, then
immediate transitions can actually be fulfilled. We will show below that
this shift is feasible. One can deviate the initial states away from the
entanglement-unentanglement boundary through appropriate constructions using
fully entangled Bell states.
\begin{equation}
\left\vert \alpha ^{\pm }\right\rangle \equiv \left\vert \alpha _{p=0}^{\pm
}\right\rangle
\end{equation}%
and
\begin{equation}
\left\vert \beta ^{\pm }\right\rangle \equiv \left\vert \beta _{p=0}^{\pm
}\right\rangle \text{.}
\end{equation}

\subsection{Recipe for immediate entanglement death-birth transition in Bell
states}

\label{sec:rl_rec}

We discuss the recipe for preparing initial states at desired quantum
configurations near the entanglement-unentanglement boundary for the
realization of ESD and ESB illustrated in Fig. \ref{fig:traj}. Consider the
initial reduced DM,%
\begin{equation}
\rho _{0}=w_{\alpha ^{+}}\left\vert \alpha ^{+}\right\rangle \left\langle
\alpha ^{+}\right\vert +w_{\alpha ^{-}}\left\vert \alpha ^{-}\right\rangle
\left\langle \alpha ^{-}\right\vert +w_{\beta ^{+}}\left\vert \beta
^{+}\right\rangle \left\langle \beta ^{+}\right\vert +w_{\beta
^{-}}\left\vert \beta ^{-}\right\rangle \left\langle \beta ^{-}\right\vert
\text{,}  \label{eq:Bell_rho0}
\end{equation}%
with the weighting $W=\left( w_{\alpha ^{+}},w_{\alpha ^{-}},w_{\beta
^{+}},w_{\beta ^{-}}\right) $ being normalized, $w_{\alpha ^{+}}+w_{\alpha
^{-}}+w_{\beta ^{+}}+w_{\beta ^{-}}=1$. Through straightforward
computations, we find that the eigenvalues of the partial transpose, $\rho
_{0}^{T_{B}}$ in (\ref{eq:PTB_rho}), are%
\begin{equation}
\lambda \in \left\{ \frac{1}{2}-w_{\alpha ^{+}},\frac{1}{2}-w_{\alpha ^{-}},%
\frac{1}{2}-w_{\beta ^{+}},\frac{1}{2}-w_{\beta ^{-}}\right\} \text{.}
\label{eq:CNE}
\end{equation}%
The negativity $\mathcal{N}$ thus aligns with the concurrence $\mathcal{C}$
in Eq. (\ref{eq:C_Jt}), with the $\gamma $ evaluated as $\gamma \in \left\{
w_{\alpha ^{+}},w_{\alpha ^{-}},w_{\beta ^{+}},w_{\beta ^{-}}\right\} $;
both monotones signify the presence of entanglement if any of the weights $%
w_{\alpha ^{+/-}}$ and $w_{\beta ^{+/-}}$exceed one-half. As a reminder for
the initial DM (\ref{eq:Bell_rho0}), note again that the CNE is readily
identifiable because $\rho _{0}^{T_{B}}$ has at most one negative eigenvalue.

\paragraph*{\textbf{Definition 2.}}

If varying the sign (including zero) of the state parameter $\varepsilon $
transforms an entangled state into a separable one (or vice versa), then $%
\varepsilon $ is called an ESP.

To construct the ESP defined above, we first focus on the case of two
nonzero assigned weights in Eq. (\ref{eq:Bell_rho0}). In this case, there is
always one negative eigenvalue in $\lambda $ when neither weight equals
one-half. Thus, the ESP can be formulated through the following weighting
assignment $\left\{ 1/2+\varepsilon /2,1/2-\varepsilon /2\right\} $.
According to Eq. (\ref{eq:CNE}), $sign(\varepsilon )\neq 0$ switches on the
entanglement [as demonstrated by the trajectories $p_{3}$ and $p_{6}$ in
Fig. \ref{fig:traj}(b)], while $sign(\varepsilon =0)\equiv 0$ switches it
off [trajectories $p_{1}$ and $p_{2}$ in Fig. \ref{fig:traj}(a)].

\paragraph*{\textbf{Definition 3.}}

An ESP $\varepsilon $ is said to be penetrable if opposite signs of $%
\varepsilon $ correspond to entangled and separable states, respectively.

Based on this penetrability definition, the previous case involving two
nonzero weights yields only impenetrable ESPs, since both $sign(\varepsilon
)>0$ and $sign(\varepsilon )<0$ correspond to entangled states. To construct
a penetrable ESP, we consider the case involving more than two Bell states.
In such scenarios, entanglement can be switched using $1/2+\varepsilon /2$
as one of the weights. For example, the weighting $\left\{ 1/2+\varepsilon
/2,1/4-\varepsilon /4,1/4-\varepsilon /4\right\} $ in Eq. (\ref{eq:Bell_rho0}%
) includes three Bell states, while $\left\{ 1/2+\varepsilon
/2,1/4-\varepsilon /4,1/4-\varepsilon /4\right\} $ includes four. Thus,
opposite signs of $\varepsilon $ yield correspondingly entangled [$%
sign(\varepsilon )>0$] and unentangled [$sign(\varepsilon )<0$] states; that
is, tuning $\varepsilon \rightarrow -\varepsilon $ across zero permits the
state to penetrate the entanglement-unentanglement boundary. Then, with
the CNE identified from our result in Eq. (\ref{eq:CNE}), the following can
be concluded.

\paragraph*{\textbf{Proposition 1.}}

If a state of the form (\ref{eq:Bell_rho0}) is expanded over more than two
Bell state components, then a penetrable ESP exists.

By preparing the state with at least one component weighted near one-half,
we have already explicitly demonstrated the proposition. To take advantage
of the penetrability for observing ESD and ESB, we notice the following
proposition.

\paragraph*{\textbf{Proposition 2.}}

If a tendency toward entanglement or unentanglement exists and persists
under a sign change of a penetrable ESP, then either ESD or ESB can be
observed.

This statement can be realized by examining the trajectories near the
entanglement-unentanglement boundary, as shown in Fig. \ref{fig:traj}. For
example, if one prepares the initial entangled state $p_{3}$, which exhibits
a tendency toward increased entanglement over time, then swapping the sign
of $\varepsilon $ (namely, $\varepsilon \rightarrow -\varepsilon $)
effectively yields the mapping $p_{3}\rightarrow p_{4}$, without altering
the tendency, thereby enabling the occurrence of ESB. Conversely, if one
prepares the initial separable state $p_{5}$, which exhibits a tendency
toward decreased entanglement, then the sign swap yields $p_{5}\rightarrow
p_{6}$, resulting in ESD. Moreover, if the $dt^{2}$ symmetry is also
satisfied, the whole (including $t<0$) trajectory, $p_{1,2,\cdots ,6}$ in
Fig. \ref{fig:traj}, is valid; one thus attains the death-to-birth
transition $p_{4}$ and birth-to-death transition $p_{6}$.

Accordingly, the recipe for the TFD proceeds as follows: (\textit{i})
Prepare initial states with the weighting $W$, as described above, using a
trial sign for $\varepsilon $. (\textit{ii}) If ESD and ESB do not take
place in the near future, swap the sign $\varepsilon \rightarrow
-\varepsilon $. (\textit{iii}) Check if the system has any of the symmetries
depicted in (\ref{eq:sys_UJmt_UmJt}), (\ref{eq:sys_drhobar_dtbar}), and (\ref%
{eq:sys_NJt_NJmdt}); utilize (\ref{eq:sys_UJmt_UmJt}) or (\ref%
{eq:sys_drhobar_dtbar}) to convert ESD into ESB or vice versa and utilize (%
\ref{eq:sys_NJt_NJmdt}) to attain the death-birth transition. Being worth
mentioning, the recipe for realizing the ESD and ESB described above does
not rely on specific types of interacting Hamiltonians. Although when
adopting only two weighted Bell states the ESP is not penetrable (i.e.,
entanglement persists for both positive and negative $\varepsilon $, and
hence the recipe may fail), there is still a chance to witness the
nontrivial transition $p_{6}$, as illustrated below.

\begin{table*}[tbph]
\caption{Immediate evolution of characteristic negative eigenvalue $\protect%
\lambda ^{\ast }\left( dt\right) $ for different mixing weightings, $%
W_{1,2,\cdots 6}$ incorporating two Bell states, $W_{7,8,\cdots ,10}$ three
Bell states, and $W_{11,12,\cdots ,14}$ four Bell states in Eq. (\protect\ref%
{eq:Bell_rho0}). The environment has $S^{C}=1/2$ initially at spin-$z$ up.
The chosen sign of $\protect\varepsilon $ is listed in the second column,
with the corresponding trajectory (see Fig. \protect\ref{fig:traj})
approaching the entanglement-unentanglement boundary shown in the last
column. Here for $W_{6}$, specifically, the leading term in $-O\left( dt^{4}/%
\protect\varepsilon \right) $ is $-dt^{4}J_{x}^{2}J_{y}^{2}\left( 1+\protect%
\varepsilon \right) \left( 3+7\protect\varepsilon \right) /\left( 12\protect%
\varepsilon \right) $ and $O$ $\left( dt^{4}/\protect\varepsilon \right) $
is $dt^{4}J_{x}^{2}J_{y}^{2}\left( 1+\protect\varepsilon \right) ^{2}/\left(
4\protect\varepsilon \right) $ .}
\label{tab:etABC}\centering
%Put tabular Here
\begin{tabular}{c|c|c|c}
\toprule$%
\begin{array}{c}
\text{Weighting} \\
\left( w_{\alpha ^{+},}w_{\alpha ^{-},}w_{\beta ^{+},}w_{\beta ^{-}}\right)%
\end{array}%
$ & $%
\begin{array}{c}
\text{Sign}(\varepsilon)%
\end{array}%
$ & $%
\begin{array}{c}
\text{Characteristic} \\
\text{negative eigenvalue }\lambda ^{\ast }\left( dt\right)%
\end{array}%
$ & $%
\begin{array}{c}
\text{Tra-} \\
\text{jectory}%
\end{array}%
$ \\ \hline
$W_{1}=\left( \frac{1+\varepsilon }{2},\frac{1-\varepsilon }{2},0,0\right) $
& $+$ & $\frac{-\varepsilon }{2}+dt^{2}\frac{J_{x}^{2}\left( 1+\varepsilon
\right) }{2}$ & $p_{6}$ \\ \hline
$W_{2}=\left( \frac{1+\varepsilon }{2},0,\frac{1-\varepsilon }{2},0\right) $
& $+$ & $\frac{-\varepsilon }{2}+dt^{2}J_{x}^{2}\varepsilon $ & $p_{6}$ \\
\hline
$W_{3}=\left( \frac{1+\varepsilon }{2},0,0,\frac{1-\varepsilon }{2}\right) $
& $+$ & $\frac{-\varepsilon }{2}+dt^{2}\frac{J_{x}^{2}\left( 1+\varepsilon
\right) }{2}$ & $p_{6}$ \\ \hline
$W_{4}=\left( 0,\frac{1+\varepsilon }{2},\frac{1-\varepsilon }{2},0\right) $
& $+$ & $\frac{-\varepsilon }{2}+dt^{2}J_{y}^{2}\varepsilon $ & $p_{6}$ \\
\hline
$W_{5}=\left( 0,\frac{1+\varepsilon }{2},0,\frac{1-\varepsilon }{2}\right) $
& $+$ & $\frac{-\varepsilon }{2}+dt^{2}\frac{J_{y}^{2}\left( 1+\varepsilon
\right) }{2}$ & $p_{6}$ \\ \hline
$W_{6}=\left( 0,0,\frac{1+\varepsilon }{2},\frac{1-\varepsilon }{2}\right) $
& $\pm $ & $%
\begin{array}{c}
\frac{-\varepsilon }{2}+dt^{2}\frac{\left( 1+\varepsilon \right) \left(
J_{x}^{2}+J_{y}^{2}\right) }{2}-O\left( \frac{dt^{4}}{\varepsilon }\right)
\text{, if }\varepsilon >0 \\
\frac{\varepsilon }{2}+O\left( \frac{dt^{4}}{\varepsilon }\right) \text{, if
}\varepsilon <0%
\end{array}%
$ & $p_{3}$ \\ \hline
$W_{7}=\left( \frac{1+\varepsilon }{2},\frac{1-\varepsilon }{4},\frac{%
1-\varepsilon }{4},0\right) $ & $+$ & $\frac{-\varepsilon }{2}+dt^{2}\frac{%
J_{x}^{2}\left( 1+3\varepsilon \right) }{4}$ & $p_{6}$ \\ \hline
$W_{8}=\left( 0,\frac{1+\varepsilon }{2},\frac{1-\varepsilon }{4},\frac{%
1-\varepsilon }{4}\right) $ & $+$ & $\frac{-\varepsilon }{2}+dt^{2}\frac{%
J_{y}^{2}\left( 1+3\varepsilon \right) }{4}$ & $p_{6}$ \\ \hline
$W_{9}=\left( \frac{1-\varepsilon }{4},0,\frac{1+\varepsilon }{2},\frac{%
1-\varepsilon }{4}\right) $ & $+$ & $\frac{-\varepsilon }{2}+dt^{2}\left[
\frac{J_{x}^{2}\left( 1+3\varepsilon \right) }{4}+\frac{J_{y}^{2}\left(
1+\varepsilon \right) }{2}\right] $ & $p_{6}$ \\ \hline
$W_{10}=\left( \frac{1-\varepsilon }{4},\frac{1-\varepsilon }{4},0,\frac{%
1+\varepsilon }{2}\right) $ & $-$ & $\frac{-\varepsilon }{2}-dt^{4}\frac{%
J_{x}^{2}J_{y}^{2}\left( -1+\varepsilon \right) ^{2}}{8\left( 1+\varepsilon
\right) }$ & $p_{4}$ \\ \hline
$W_{11}=\left( \frac{1+\varepsilon }{2},\frac{1-\varepsilon }{6},\frac{%
1-\varepsilon }{6},\frac{1-\varepsilon }{6}\right) $ & $+$ & $\frac{%
-\varepsilon }{2}+dt^{2}\frac{J_{x}^{2}\left( 1+2\varepsilon \right) }{3}$ &
$p_{6}$ \\ \hline
$W_{12}=\left( \frac{1-\varepsilon }{6},\frac{1+\varepsilon }{2},\frac{%
1-\varepsilon }{6},\frac{1-\varepsilon }{6}\right) $ & $+$ & $\frac{%
-\varepsilon }{2}+dt^{2}\frac{J_{y}^{2}\left( 1+2\varepsilon \right) }{3}$ &
$p_{6}$ \\ \hline
$W_{13}=\left( \frac{1-\varepsilon }{6},\frac{1-\varepsilon }{6},\frac{%
1+\varepsilon }{2},\frac{1-\varepsilon }{6}\right) $ & $+$ & $\frac{%
-\varepsilon }{2}+dt^{2}\frac{\left( J_{x}^{2}+J_{y}^{2}\right) \left(
1+2\varepsilon \right) }{3}$ & $p_{6}$ \\ \hline
$W_{14}=\left( \frac{1-\varepsilon }{6},\frac{1-\varepsilon }{6},\frac{%
1-\varepsilon }{6},\frac{1+\varepsilon }{2}\right) $ & $-$ & $\frac{%
-\varepsilon }{2}-dt^{4}\frac{J_{x}^{2}J_{y}^{2}\left( -1+\varepsilon
\right) ^{2}}{3+6\varepsilon }$ & $p_{4}$ \\
\bottomrule
\end{tabular}%
\end{table*}

\subsection{Immediate transition with classical weighting (mixed states)}

\label{sec:rl_cla}

\begin{figure}[tbph]
%\begin{figure*} for two-column figure
\centerline{\psfig{file=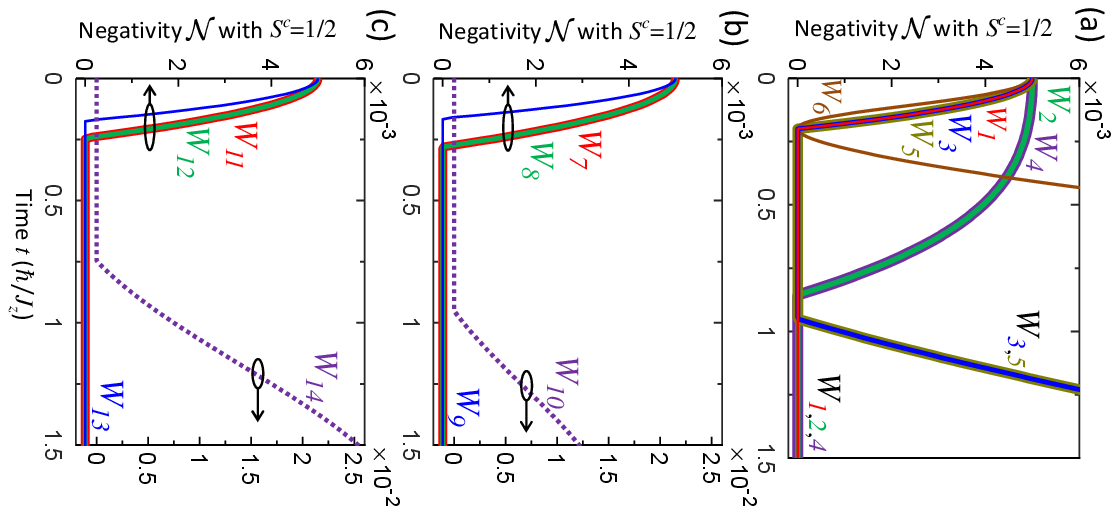,scale=0.92,angle=90,clip=true,
trim=9mm 173mm 0mm 0mm}}%For angle=90(270), trim=b(t),r(l),t(b),l(r)
%t=top,b=bottom,r=right,l=left
%Below is for inserting another figure.
%\centerline{\psfig{file=figX.eps,scale=0.33,angle=270,clip=true,trim=0mm 0mm 0mm 0mm}}
\caption{Negativity development in mixed states with $\left\vert \protect%
\varepsilon \right\vert =10^{-2}$ and ferromagnetic $\vec{J}=\left(
-0.5,-0.5,-1\right) $. The environment carries quantum degrees of freedom $%
2S^{C}+1=2$ initially aligned with spin-$z$ up. The mixing weightings $W$s
in Table \protect\ref{tab:etABC} are adopted. In (a), (b), and (c), two,
three, and four Bell states are used, respectively. Positive (solid lines,
left axis) entanglement switch parameters $\protect\varepsilon >0$ are
chosen for all $W$s, except for $W_{10}$ and $W_{14}$ (dotted lines, right
axis). The near-past and near-future negativity obeys $\mathcal{N}(\vec{J}%
,dt)=\mathcal{N}(\vec{J},-dt)$, although the $t\lesssim 0$ region is not
shown.}
\label{fig:mixNEdt}
\end{figure}

To demonstrate the recipe for mixed states $Tr\left( \varrho _{0}^{2}\right)
<1$, we present analytic results based on (\ref{eq:dtexp}) and numerical
simulations on (\ref{eq:emo_rho}). We choose $S^{C}=1/2$ at spin up $%
\left\vert \uparrow \right\rangle $ and the initial DM%
\begin{equation}
\varrho _{0}=\left\vert \uparrow \right\rangle \left\langle \uparrow
\right\vert \otimes \sum_{i=\alpha ^{+},\alpha ^{-},\beta ^{+},\beta
^{-}}w_{i}\left\vert i\right\rangle \left\langle i\right\vert
\end{equation}%
of the classical mixing weights (\ref{eq:Wweight}) specifically listed in
Table \ref{tab:etABC}, where analytic expressions of entanglement dynamics
are provided. Table \ref{tab:etABC} expands $\lambda ^{\ast }\left(
dt\right) $ up to the first two leading orders using two Bell states in the
weightings $W_{1}$, $W_{2}$, $\cdots $, $W_{5}$, and $W_{6}$, using three
Bell states in $W_{7}$, $W_{8}$, $W_{9}$, and $W_{10}$, while using four
Bell states in $W_{11}$, $W_{12}$, $W_{13}$, and $W_{14}$. The desired sign
of $\varepsilon $ is selected to achieve immediate ESD or ESB, and
consequently, a nontrivial TFD due to the $dt^{2}$ symmetry (see Table \ref%
{tab:etABC}). Furthermore, for our spin-star model, Table \ref{tab:etABC}
indicates that nontrivial TFD can all be identified except for weighting $%
W_{6}$, where $\varepsilon $ is not penetrable. For weighting $W_{6}$ of
small $\varepsilon $, the CNE $\lambda ^{\ast }\left( dt\right) $ stays in
trajectory $p_{3}$ for both signs of $\varepsilon $, where $\lambda ^{\ast
}\left( dt\right) $ is predominantly governed by $O\left( dt^{4}/\varepsilon
\right) $ if $\varepsilon <0$ and $-O\left( dt^{4}/\varepsilon \right) $ if $%
\varepsilon >0$ (as given in Table \ref{tab:etABC}), while in trajectory $%
p_{2}$ if $\varepsilon =0$. For $W_{10}$ and $W_{14}$, the higher-order $%
dt^{4}$ term leads to ESB if $\varepsilon <0$ is chosen, corresponding to
trajectory $p_{6}$ of TFD. Interestingly, $J_{z}$ is absent in determining
the immediate dynamics of entanglement, similar to the case of direct
exchange in Eq. (\ref{eq:Cdir}).

The numerical simulation for the negativity is shown in Fig. \ref%
{fig:mixNEdt}, where we adopt $\vec{J}=\left( -0.5,-0.5,-1\right) $ and $%
\left\vert \varepsilon \right\vert =10^{-2}$. The result coincides with
Table \ref{tab:etABC}. In Fig. \ref{fig:mixNEdt}, the positive (solid lines,
left axis) ESP $\varepsilon $ for all $W$s is selected, except for $W_{10}$
and $W_{14}$, where negative $\varepsilon <0$ (dashed lines, right axis)
offers ESB depicted by trajectory $p_{4}$. Note that Fig. \ref{fig:mixNEdt}
does not plot the negativity in the near-past $t\lesssim 0$ region, which is
actually identical to the negativity in the near-future $t\gtrsim 0$ region,
attributed to the $dt^{2}$ symmetry provided in Table \ref{tab:etABC}. In
other words, the immediate TFD is obtained with $p_{4}$ (death-to-birth) and
$p_{6}$ (birth-to-death) by further including the near-past entanglement
dynamics. For weighting $W_{6}$ in Fig. \ref{fig:mixNEdt}, the negativity $%
\mathcal{N}(\vec{J},dt)$ first decreases due to $dt^{2}$ term and then
increases due to the $dt^{4}/\varepsilon $ term, in line with the expression
in Table \ref{tab:etABC}. Moreover, as Table \ref{tab:etABC} implies, the
immediate $\mathcal{N}(\vec{J},dt)$ remains unaffected by the strength of $%
J_{z}$ and the sign of $\vec{J}$ in our spin-star model.

\subsection{Immediate transition with quantum weighting (pure states)}

\label{sec:rl_qua}

\begin{figure}[tbph]
%\begin{figure*} for two-column figure
\centerline{\psfig{file=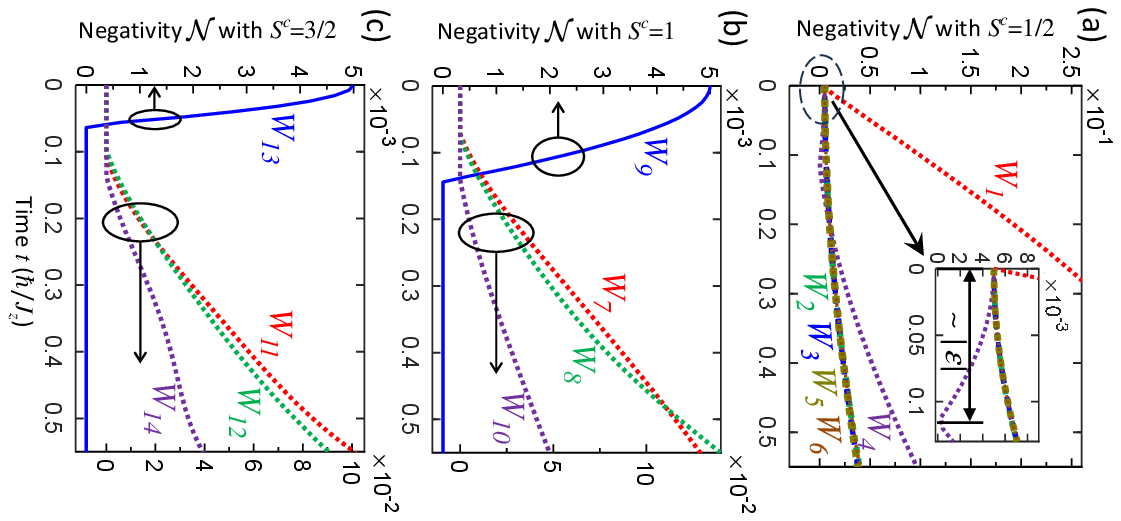,scale=0.92,angle=90,clip=true,
trim=9mm 170mm 0mm 0mm}}%For angle=90(270), trim=b(t),r(l),t(b),l(r)
%t=top,b=bottom,r=right,l=left
%Below is for inserting another figure.
%\centerline{\psfig{file=figX.eps,scale=0.33,angle=270,clip=true,trim=0mm 0mm 0mm 0mm}}
\caption{Negativity development in pure states with $\left\vert \protect%
\varepsilon \right\vert =10^{-2}$ and ferromagnetic $\vec{J}=\left(
-0.5,-0.5,-1\right) $. The environment possesses quantum degrees of freedom
expanded by $2S^{C}+1$ basis states with (a) $S^{C}=1/2$, (b) $S^{C}=1$, and
(c) $S^{C}=3/2$ associating with two, three, and four Bell states,
respectively. The inset in (a) provides a zoomed-in view of $W_{4}$ around $%
t=0$, with the local minimum of negativity occurring at the time $t\approx
0.11$ quantified by $\left\vert \protect\varepsilon \right\vert $. The
quantum weightings $W$s are assigned through Eq. (\protect\ref{eq:purePsi_0}%
) according to Table \protect\ref{tab:etABC}. Solid (dotted) lines for the
left (right) axis represent the adopted positive (negative) entanglement
switch parameters $\protect\varepsilon >0$ ($\protect\varepsilon <0$). The
near-past negativity $\mathcal{N}(\vec{J},-dt)$ is identical to $\mathcal{N}(%
\vec{J},dt)$.}
\label{fig:pureNEdt}
\end{figure}

Following the recipe outlined in Sec. \ref{sec:rl_rec}, we demonstrate the
existence of the immediate TFD for pure states $Tr\left( \varrho
_{0}^{2}\right) =1$, with $\varrho _{0}=\left\vert \psi _{0}\right\rangle
\left\langle \psi _{0}\right\vert $ in the form (\ref{eq:Qweight}), provided
$\varepsilon $ is penetrable. Specifically, consider initial%
\begin{equation}
\left\vert \psi _{0}\right\rangle =\sum_{\substack{ i=\alpha ^{+},\alpha
^{-},\beta ^{+},\beta ^{-}  \\ w_{i}\neq 0}}\sqrt{w_{i}}\left\vert
i^{C}\right\rangle \otimes \left\vert i\right\rangle \text{.}
\label{eq:purePsi_0}
\end{equation}%
where the summation accounts for only the nonzero weights $w_{i}\neq 0$
chosen from Table \ref{tab:etABC}. The $\left\vert i^{C}\right\rangle $
loops over the spin-$z$ eigenspinors $\left\vert m\right\rangle $ of $C$ in
descending order, $\left\vert m=S^{C}\right\rangle $, $\left\vert
S^{C}-1\right\rangle $, $\cdots $, and $\left\vert -S^{C}\right\rangle $;
the $S^{C}=1/2$ is set to match the number of nonzero $w_{i}$ in each of the
weightings $W_{1,2,\cdots 6}$, $S^{C}=1$ in $W_{7,8,\cdots 10}$, and $%
S^{C}=3/2$ in $W_{11,12,\cdots 14}$. For example, according to Table \ref%
{tab:etABC}, for $W_{9}$, we assign
\begin{eqnarray}
\left\vert \psi _{0}\right\rangle &=&\sqrt{\frac{1-\varepsilon }{4}}%
\left\vert m=1\right\rangle \otimes \left\vert \alpha ^{+}\right\rangle
\notag \\
&&+\sqrt{\frac{1+\varepsilon }{2}}\left\vert m=0\right\rangle \otimes
\left\vert \beta ^{+}\right\rangle  \notag \\
&&+\sqrt{\frac{1-\varepsilon }{4}}\left\vert m=-1\right\rangle \otimes
\left\vert \beta ^{-}\right\rangle ,
\end{eqnarray}%
while for $W_{13}$, we assign
\begin{eqnarray}
\left\vert \psi _{0}\right\rangle &=&\sqrt{\frac{1-\varepsilon }{6}}%
\left\vert m=3/2\right\rangle \otimes \left\vert \alpha ^{+}\right\rangle
\notag \\
&&+\sqrt{\frac{1-\varepsilon }{6}}\left\vert m=1/2\right\rangle \otimes
\left\vert \alpha ^{-}\right\rangle  \notag \\
&&+\sqrt{\frac{1+\varepsilon }{2}}\left\vert m=-1/2\right\rangle \otimes
\left\vert \beta ^{+}\right\rangle  \notag \\
&&+\sqrt{\frac{1-\varepsilon }{6}}\left\vert m=-3/2\right\rangle \otimes
\left\vert \beta ^{-}\right\rangle .
\end{eqnarray}%
Because $\left\vert m\right\rangle \otimes \left\vert i\right\rangle $
constitutes an orthonormal basis, tracing out $C$ in fact yields the same
initial reduced DM, Eq. (\ref{eq:Bell_rho0}). Consequently, the previously
established recipe remains applicable to the pure states (\ref{eq:purePsi_0}%
) by associating different $\left\vert i^{C}\right\rangle $ with different $%
\left\vert m\right\rangle $, which avoids cross term like $\left\vert \alpha
^{\pm }\right\rangle \left\langle \beta ^{\pm }\right\vert $ in $\rho _{0}$.
Furthermore, employing again expansion (\ref{eq:dtexp}) up to the second
term, it can be shown that the pure states with the weights listed in Table %
\ref{tab:etABC} satisfy $dt^{2}$ symmetry.

Figure \ref{fig:pureNEdt} plots the negativity for pure states with $\vec{J}%
=\left( -0.5,-0.5,-1\right) $ and $\left\vert \varepsilon \right\vert
=10^{-2}$. In the weightings $W_{1}$, $W_{2}$, $\cdots $, $W_{5}$, and $%
W_{6} $, where only two Bell states are incorporated in $\varrho _{0}$, the
lack of penetrability of $\varepsilon $ impedes the occurrence of ESD and
ESB in Fig. \ref{fig:pureNEdt}(a). Neither sign of $\varepsilon $ gives rise
to a nontrivial transition such as $p_{6}$. For example, $W_{4}$ in \ref%
{fig:pureNEdt}(a) with $\varepsilon <0$ encounters a local minimum of
negativity around $t\approx 0.11$. As $\varepsilon $ is increased, at $%
\varepsilon =0$ this local minimum is located exactly at $t=0$ and becomes
zero, indicating the absence of entanglement (trajectory $p_{2}$). With
further increases to $\varepsilon >0$, entanglement reappears at $t=0$
(trajectory $p_{3}$). On the contrary, when more than two Bell states are
utilized in $\varrho _{0}$, the penetrability of $\varepsilon $ expectedly
enables the TFD, as demonstrated in Figs. \ref{fig:pureNEdt}(b) with three
Bell states and \ref{fig:pureNEdt}(c) with four Bell states. The weightings $%
W_{9}$ and $W_{13}$, with positive $\varepsilon >0$ (solid lines, left
axis), render the entanglement birth-to-death transition (trajectory $p_{6}$%
), while other weightings with negative $\varepsilon <0$ (dotted lines,
right axis) result in the entanglement death-to-birth transition (trajectory
$p_{4}$).

\section{Summary}

\label{sec:summ}

In summary, to identify the ESD, ESB, and TFD in Fig. \ref{fig:sche}, we
propose a recipe for moving initial states around the
entanglement-unentanglement boundary by introducing the ESP $\varepsilon $.
The spin-star model of two qubits serves as an example to illustrate this
recipe. To appreciate that the TFD is precious and nontrivial, we first
examined the fully unentangled and individually pure states in Table \ref%
{tab:sepABC} and Fig. \ref{fig:longt}, where all \emph{immediate}
transitions, characterized by the nonzero CNE $\lambda ^{\ast }\left(
dt\right) $, exhibit \emph{zero} duration (trajectory $p_{2}$ in Fig. \ref%
{fig:traj}). When $\rho ^{C}$ is a mixed state in a similar fully unentangled
configuration, the $\lambda ^{\ast }\left( dt\right) $ in Eq. (\ref%
{eq:lamda_star_Csepmix}), depicts again trivial transition (trajectory $%
p_{2} $). When $C$ state and $A$-$B$ state are both pure, where $A$-$B$ is
described by one Bell state unentangled with $C$, the leading term of finite
constant in Eqs. (\ref{eq:lambda_star_alpha}) and (\ref{eq:lambda_star_beta}%
) even implies no immediate transitions.

Viewing the local time-even ($dt^{2}$) symmetry Eq. (\ref{eq:sys_NJt_NJmdt})
of $p_{2}$ as difficult to avoid, and examining all other trajectories in
Fig. \ref{fig:traj}, we then realized that we can leverage this symmetry
instead of avoiding it to induce the ESD and ESB. Such a transition is
achieved by shifting the symmetry point away from the boundary via $%
\varepsilon $, encoded in the weights of the Bell states Eq. (\ref%
{eq:Bell_rho0}). When more than two Bell states are adopted (see $W_{7}$, $%
W_{8}$, $\cdots $, and $W_{14}$ in Table \ref{tab:etABC}), the initial state
is penetrable through the boundary controlled by $sign(\varepsilon )$. The
penetrability enables the immediate occurrence of the ESD and ESB, and when
the $dt^{2}$ symmetry is satisfied, it ensures the entanglement death-birth
transition for both mixed $\varrho _{0}$ in Figs. \ref{fig:mixNEdt}(b) and %
\ref{fig:mixNEdt}(c) and pure $\varrho _{0}$ in Figs. \ref{fig:pureNEdt}(b)
and \ref{fig:pureNEdt}(c). Other symmetries characterized by Eq. (\ref%
{eq:sys_UJmt_UmJt}) and Eq. (\ref{eq:sys_drhobar_dtbar}) allow conversion
between ESD and ESB, and vice versa.

The recipe with the swap $\varepsilon \rightarrow -\varepsilon $ works
generally in any interactions, under the conditions that (\textit{i}) $%
\varepsilon $ is penetrable, and (\textit{ii}) sufficiently small $%
\varepsilon $ will only displace the trajectories without altering the
entanglement tendency. Here, both conditions hold when more than two Bell
states are employed. However, the penetrability is lost when only two Bell
states are used to construct $\varrho _{0}$. In such cases, the latter
condition may still be satisfied under specific configurations---For
example, see $\lambda ^{\ast }(dt)$ with weightings $W_{1}$, $W_{3}$, and $%
W_{5}$ in Table~\ref{tab:etABC}, corresponding to $p_{6}$ [as also shown in
Fig.~\ref{fig:mixNEdt}(a)].

In the case of pure\emph{\ }$\varrho $ with $S^{C}=1/2$ [solid lines in Fig. \ref%
{fig:longt} and dashed lines in Fig. \ref{fig:pureNEdt}(a)], all transitions
remain trivial, and no ESD and ESB are obtained with our recipe. We thus
conjecture that, to obtain the TFD, ESD, or ESB for a \emph{pure} $\varrho $%
, the environment must possess quantum degrees of freedom greater than those
of a qubit, namely, $2S^{C}+1>2$. Nonetheless, extending this conjecture
beyond the two Bell states used here would require a rigorous proof, which
is beyond the scope of this paper. Furthermore, a phase prefactor $\exp
\left( i\varphi _{i}\right) $ can be introduced in the construction Eq. (\ref%
{eq:purePsi_0}), which leaves $\rho _{0}$ unchanged; thus, the prefactor is
not relevant to the overall recipe. The phase $\varphi _{i}$ was neglected
in this paper, as discussing the effects from it would divert the focus
herein.

As a remark, the derivations leading to the results in Tables \ref%
{tab:sepABC} and \ref{tab:etABC}, though lengthy, are based on
straightforward standard matrix algebra. Readers interested in the explicit
computational steps may find it helpful to use symbolic computation software
such as \textit{Mathematica}, which can readily perform and display the full
calculations.

Finally, we emphasize again that the proposed recipe relies primarily on the
penetrability of the ESP, which is encoded solely through the initial
states. To the best of our knowledge, the ESP has not been introduced or
explored in the existing literature. Hence, a key advantage of this approach
is that it does not require any specific type of interaction. Furthermore,
the recipe is systematic, making it broadly applicable. In particular, by
leveraging both the controllable duration of entanglement and the immediacy
of its onset, the recipe facilitates the engineering and practical
application of entanglement dynamics. Moreover, the introduced
ESP---especially the notion of penetrability---opens new avenues for
exploring the fundamental physics of entanglement kinetics. Importantly, the
applicability of the ESP is not limited to two-level systems; the concept
can be naturally generalized beyond qubits to qudits, thereby extending its
relevance to a broader class of quantum systems. Whether ESD or
ESB occurs can be reduced to the
question of whether a penetrable ESP exists.

\begin{acknowledgments}
One of the authors, S.-H.C. thanks Shih-Jye Sun, Ming-Chien Hsu, and
Ching-Ray Chang for valuable discussions. S.-G. Tan is supported by the
National Science and Technology Council of Taiwan under Grant. No:
113-2112-M-034-002.
\end{acknowledgments}

%\bibliographystyle{apsrev}
%\bibliography{acompat,2025shc}
%
%Then, when submitting, copy and paste from the .bbl file into below
%
%\begin{thebibliography}{99}
%\bibitem{cramer} R. Cramer, R. Gennaro, and B. Schoenmakers, \emph{A Secure
%and Optimally Efficient Multi-authority Election Scheme,} in Advances in
%Cryptology (EUROCRYPT , 97), vol. 1233 of Lecture Notes in Computer Science,
%pp. 103- 118, Springer, New York, NY, USA, 1997.
%
%\bibitem{gentry} C. Gentry. \emph{A Fully Homomorphic Encryption Scheme}.
%PhD thesis, Stanford University, 2009. http://crypto.stanford.edu/craig.
%\end{thebibliography}

\newif\ifabfull\abfulltrue
\providecommand{\newblock}{}

\end{document}